\documentclass{nic-series}

\begin{document} 

\title{Multi-overlap simulations of spin glasses}

\author{Wolfhard Janke \inst{1} \and Bernd A. Berg \inst{2} \and 
        Alain Billoire \inst{3}}

\institute{Institut f\"ur Theoretische Physik, Universit\"at Leipzig,
         04109 Leipzig, Germany\\
         \email{wolfhard.janke@itp.uni-leipzig.de}
         \and
         Dept. of Physics, The Florida State University, Tallahassee,
         FL~32306, USA\\
         \email{berg@hep.fsu.edu}
         \and
         CEA/Saclay, Service de Physique Th\'eorique, 91191 Gif-sur-Yvette,
         France\\
         \email{billoir@spht.saclay.cea.fr}
         }

\maketitle

\begin{abstracts}
We present results of recent high-statistics Monte Carlo simulations of the
Edwards-Anderson Ising spin-glass model in three and four dimensions.
The study is based on a non-Boltzmann sampling technique, the
multi-overlap algorithm which is specifically tailored for sampling
rare-event states. We thus concentrate on those properties which are difficult
to obtain with standard canonical Boltzmann sampling such as the
free-energy barriers $F^q_B$ in the probability density  $P_{\cal J}(q)$
of the Parisi overlap parameter $q$ and the behaviour of the tails of the
disorder averaged density $P(q) = [P_{\cal J}(q)]_{\rm av}$.
\end{abstracts}

\section{Introduction}

\begin{figure}[t]
\vskip 6.0truecm
\includegraphics{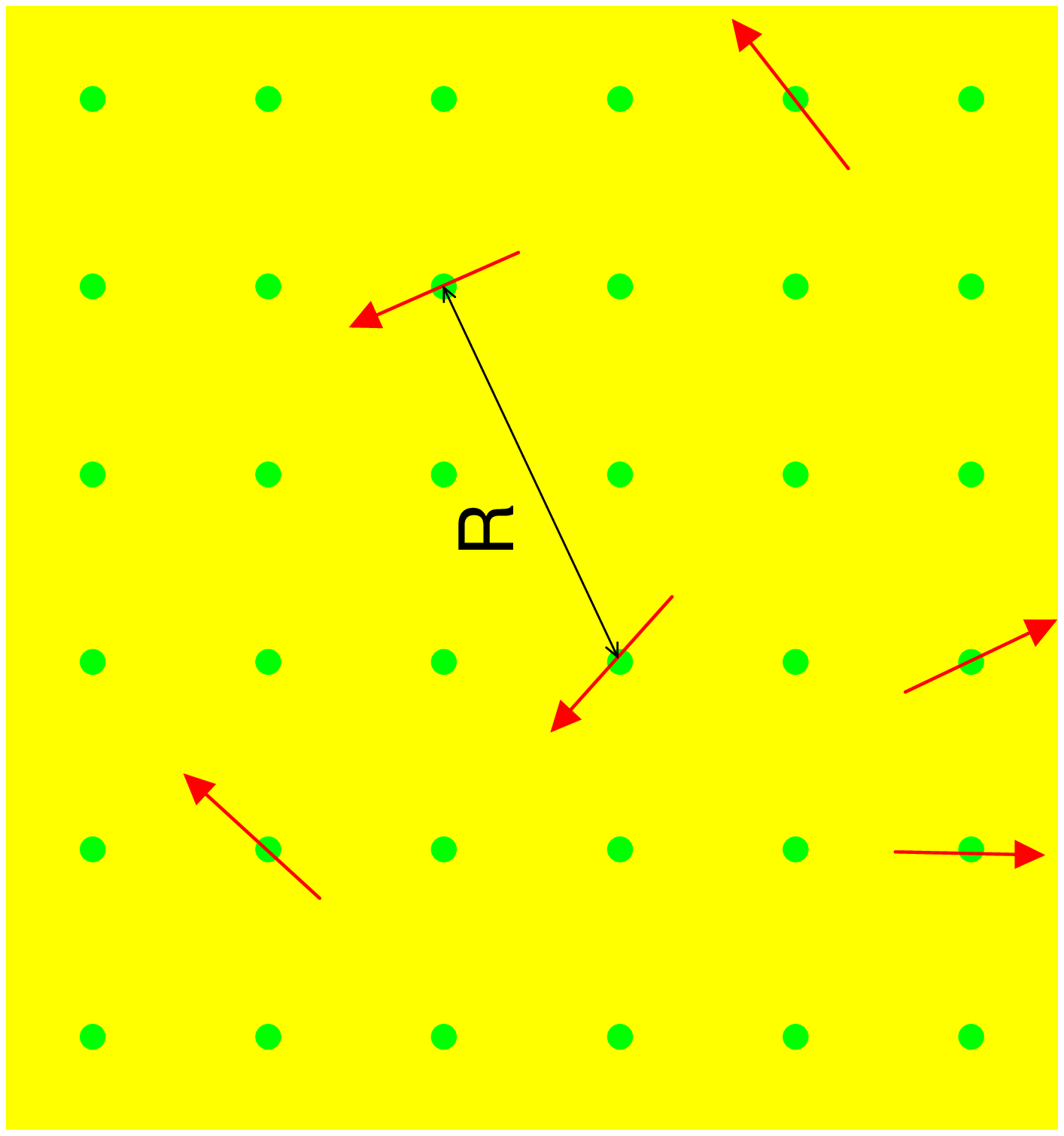}
\includegraphics{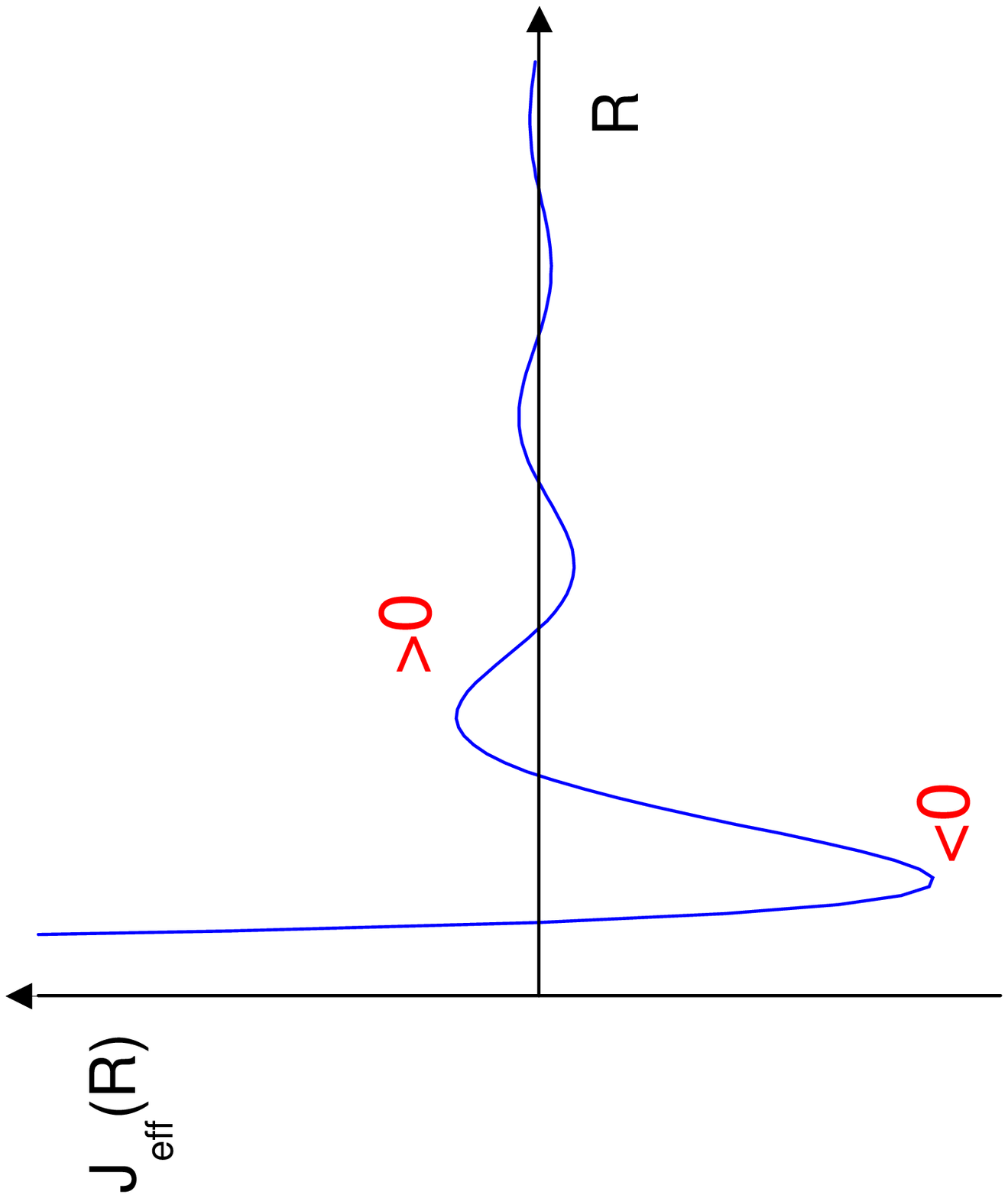}
\caption[a]{\label{fig:rkky}
The two basic ingredients for spin-glass behaviour:
{\em Randomly\/} distributed spin moments of transition metal impurities 
(e.g.\ Mn) in a noble metal host (e.g.\ Cu), and the characteristic
oscillatory form of their effective RKKY interactions with 
{\em competing\/} positive (ferromagnetic) and negative (antiferromagnetic)
regions.}
\end{figure}

\begin{figure}[-t] \begin{center}
\hspace{-0.5cm}\epsfig{figure=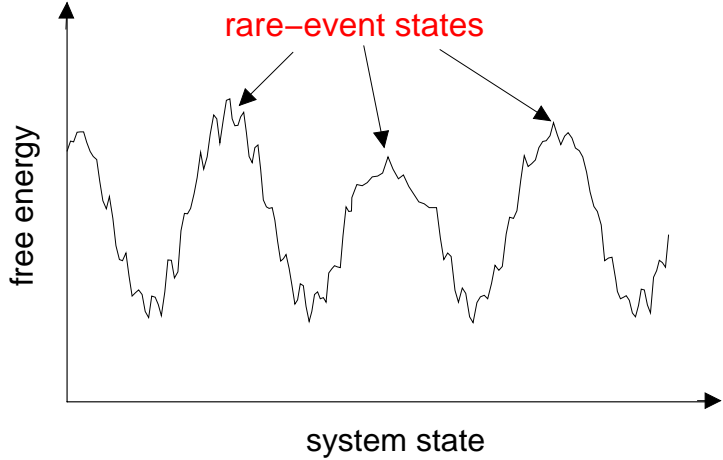,width=10cm} \vspace*{-1mm}
\end{center}
\caption[a]{\label{fig:sketch}
Typical sketch of the rugged free-energy landscape of spin glasses with
many minima separated by rare-event barriers.}
\end{figure}

A widely studied class of spin-glass materials\cite{Bi86,Me86,Fi91,Young97}
consists of dilute solutions of
magnetic transition metal impurities in noble metal 
hosts, for instance\cite{Mu82} Au-2.98\%~Mn
or\cite{Mu81} Cu-0.9\%~Mn, which is one of the best investigated metallic 
spin glasses. In these systems, the interaction between impurity moments
is caused by the polarization of the surrounding Fermi sea of the host
conduction electrons, leading to an effective interaction of the so-called
RKKY form\cite{rkky} 
\begin{equation}
J_{\rm eff}(R) = \frac{\cos(2 k_F R)}{R^3} \enspace, 
\qquad k_F R \gg 1 \enspace,
\label{eq:RKKY}
\end{equation}
where $k_F$ is the Fermi wave number. For an illustration, see Fig.~\ref{fig:rkky}.
This constitutes the two basic ingredients necessary for spin-glass behaviour, 
namely 
\begin{itemize}
\item {\em randomness\/} -- in course of the dilution process the
positions of the impurity moments are randomly distributed, and 
\item {\em competing interactions\/} -- due to the oscillations in (\ref{eq:RKKY}) as
a function of the distance $R$ between the spins some of the interactions are 
positive and some are negative. 
\end{itemize}
The competition among the 
different interactions between the moments means that no single configuration
of spins is uniquely favoured by all of the interactions, a phenomenon which 
is commonly called ``frustration''. This leads to a rugged free-energy landscape
with probable regions (low free energy) separated by rare-event states (high
free energy), illustrated in many previous articles
by vague sketches similar to our Fig.~\ref{fig:sketch}. 
Experimentally this may be inferred from the phenomenon
of aging which is typical of measurements of the remanent 
magnetization in the spin-glass phase\cite{Gr87}. 

However, despite the large amount of experimental, theoretical and simulational 
work done in the past thirty years to elucidate the spin-glass 
phase,\cite{Bi86,Me86,Fi91,Young97} the physical mechanisms underlying its peculiar
properties are not yet fully understood. To cope with the enormous complexity of
the problem various levels of simplified models have been studied theoretically. 
A simplified lattice model which reflects the two basic ingredients for
spin-glass behaviour is the Edwards-Anderson\cite{EA75}
Ising (EAI) model defined through the energy
\begin{equation} \label{eq:energy}
E = - \sum_{\langle ik \rangle} J_{ik}\, s_i s_k\ ,
\end{equation}
where the fluctuating spins $s_i$ can take the values $\pm 1$. The 
coupling constants $J_{ik}$ are quenched, {\em random\/} variables taking
positive and negative signs, thereby leading to {\em competing interactions\/}. 
In our study we worked with a bimodal
distribution, $J_{ik} = \pm 1$ with equal probabilities, but other
choices such as the Gaussian distribution, ${\cal P}(J_{ik}) \propto
\exp(-J_{ik}^2/2 \Delta^2)$ with $\Delta$ parameterizing its width,
have also been considered, in particular
in analytical work. In (\ref{eq:energy}), the lattice sum runs over 
all nearest-neighbour pairs of a $d$-dimensional (hyper-) cubic 
lattice of size $N = L^d$ with periodic boundary conditions. 

An analytically more tractable mean-field model, commonly known as
the Sherrington-Kirkpatrick\cite{SK} (SK) model, emerges when each
spin is allowed to interact with all others.
Alternatively one may
consider the mean-field treatment as an approximation which is
expected to become accurate in high dimensions\cite{Pa79}.
In physical dimensions, however, its status is still unclear 
and an alternative droplet approximation\cite{FiHu88}
has been proposed. The two treatments yield conflicting predictions which have
prompted quite a controversial discussion over many years. Numerical approaches
such as Monte Carlo (MC) simulations can, in principle, provide arbitrarily precise
results in physical dimensions. In practice, however, the simulational approach
is severely hampered by an extremely slow (pseudo-) dynamics of the stochastic
process.

To overcome the slowing-down problem various novel simulation techniques have
been devised in the past few years. While some of them only aim at improving
the (pseudo-) dynamics of the MC process, others are in addition
also well suited for a quantitative characterization of the free-energy
barriers responsible for the slowing-down problem. Among the latter category 
is the multi-overlap algorithm\cite{BeJa98} which has recently been employed in 
quite extensive MC simulations\cite{JBB98,bbj_prb00,eilat,bbj_pre01} of the EAI
spin-glass model in three and four dimensions. The purpose of this note
is to give an overview of the recent progress achieved with this method.

\section{Model Parameters and Simulation Method}

As order parameter of the EAI model one usually takes the Parisi 
overlap parameter\cite{Pa79}
\begin{equation}
q = \frac{1}{N} \sum_{i=1}^N s^{(1)}_i\, s^{(2)}_i\ ,
\label{eq:q}
\end{equation}
where the spin superscripts
label two independent (real) replicas for the same realization of randomly
chosen exchange coupling constants ${\cal J}=\{ J_{ik} \}$.
For given ${\cal J}$ the probability density of $q$ is denoted by
$P_{\cal J}(q)$, and thermodynamic expectation values are computed as
$\langle \dots \rangle_{\cal J} \equiv \sum_{\{s\}} (\dots) 
\exp(-\beta H[{\cal J}])/ \sum_{\{s\}}\exp(-\beta H[{\cal J}])$, 
where $\beta = 1/k_B T$ is the inverse
temperature in natural units. The freezing temperature is known to be at 
$\beta_c = 0.90(3)$ in 3D (Ref.~18) and at 
$\beta_c = 0.485(5)$ in 4D (Ref.~19), respectively.

On finite lattices the results necessarily depend on the randomly chosen
quenched disorder. To get a better approximation of the infinite system (apart
from special problems with so-called non-self-averaging), one performs
averages over many hundreds or even thousands of (quenched) disorder
realizations denoted by
\begin{equation}
P(q) = \left[ P_{\cal J}(q) \right]_{\rm av} =
       \frac{1}{\# {\cal J}} \sum_{\cal J} P_{\cal J}(q)\ ,\qquad
[\langle \dots \rangle_{\cal J}]_{\rm av} = 
       \frac{1}{\# {\cal J}} \sum_{\cal J} \langle \dots \rangle_{\cal J}\ ,
\label{eq:P_av}
\end{equation}
where $\# {\cal J}$ ($\rightarrow \infty$) is the number of realizations considered.
Below the freezing temperature,
in the infinite-volume limit $N \rightarrow \infty$,
a non-vanishing part of $P(q)$ between its two delta-function peaks
at $\pm q_{\rm max}$ characterizes the
mean-field picture\cite{Pa79} of spin glasses, whereas in ferromagnets as
well as in the droplet picture\cite{FiHu88} of spin glasses $P(q)$ 
exhibits only the two delta-function peaks.
Most studies so far considered mainly the averaged quantities.

\begin{figure}[t]
\vspace{14.0cm}
\includegraphics{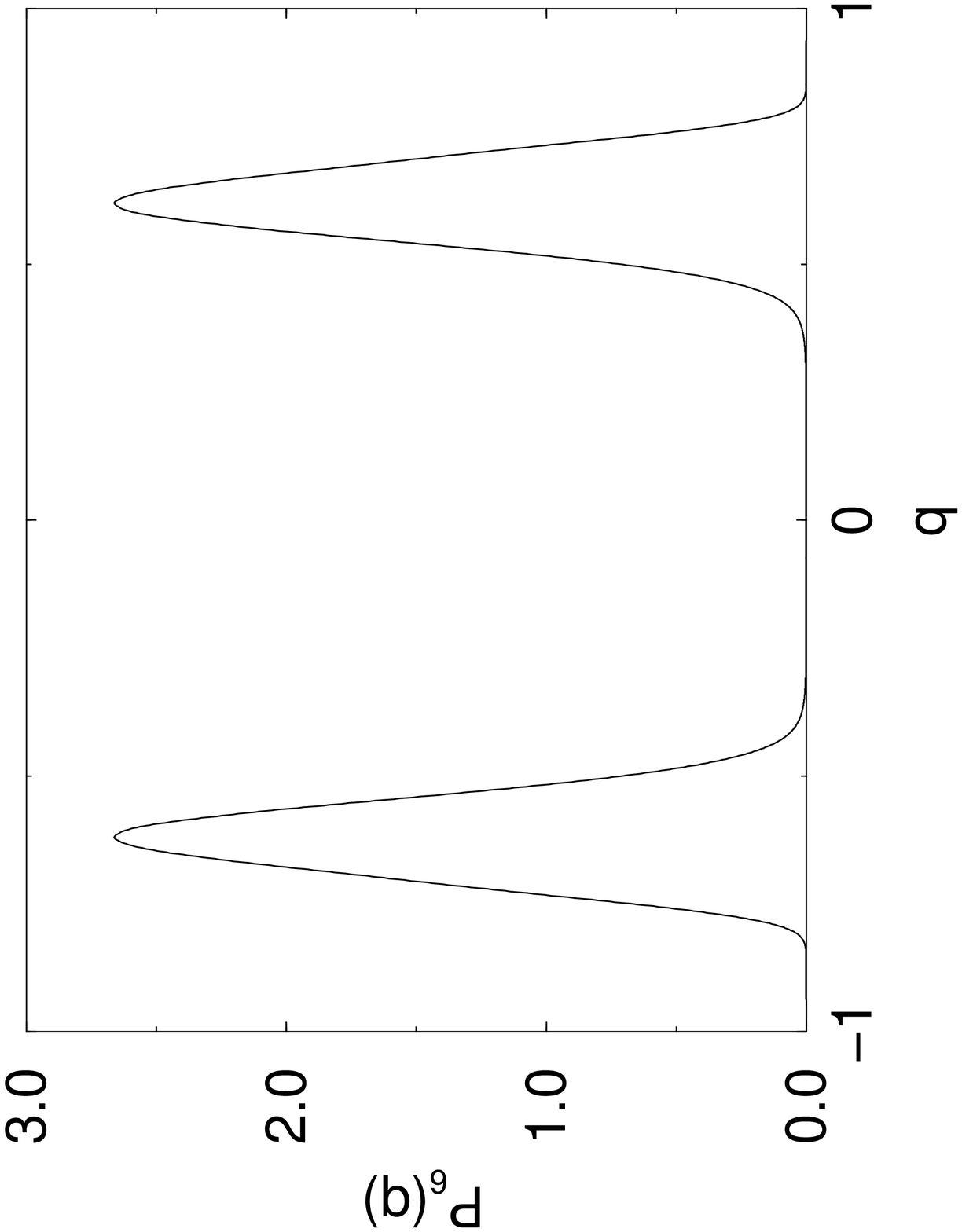}
\includegraphics{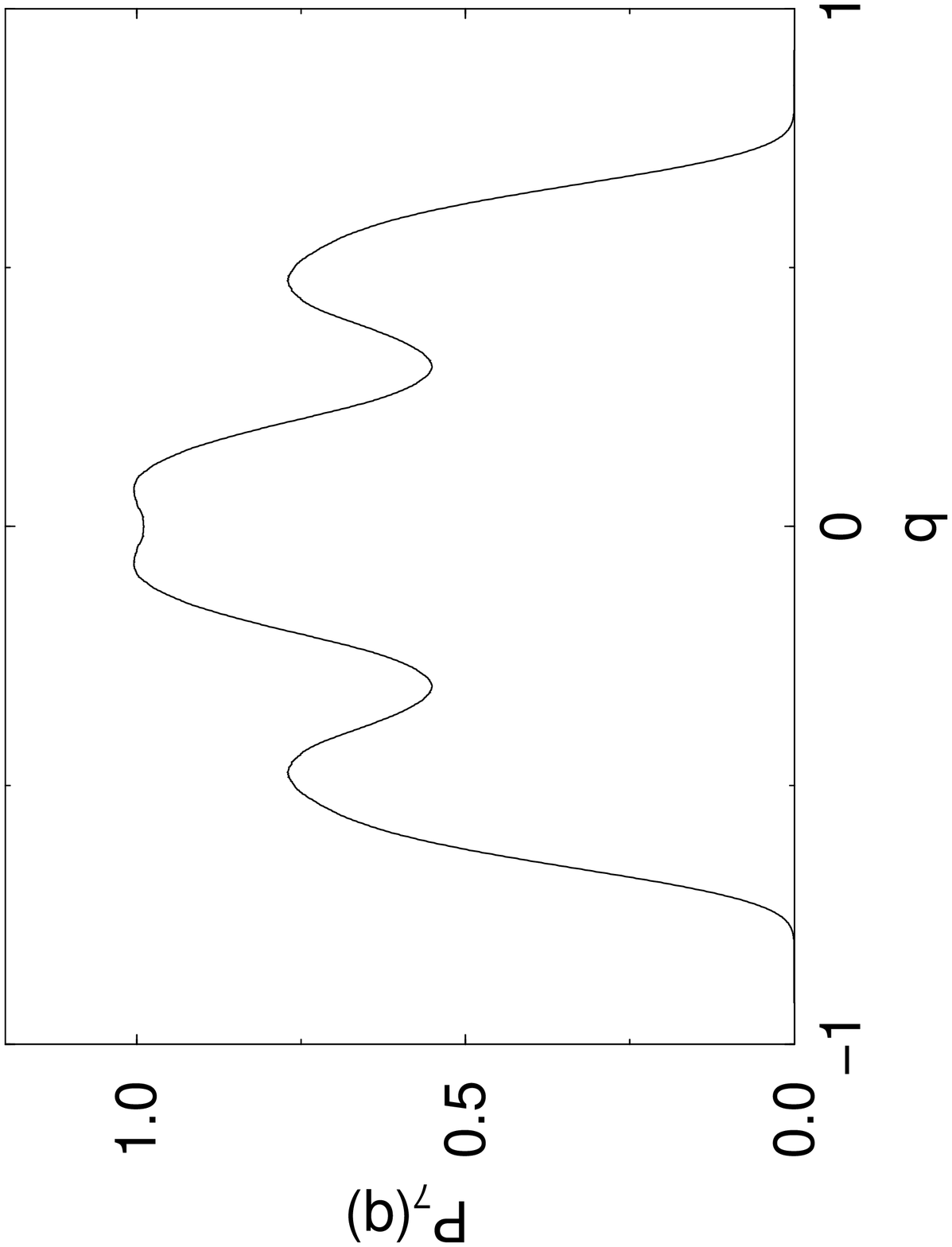}
\includegraphics{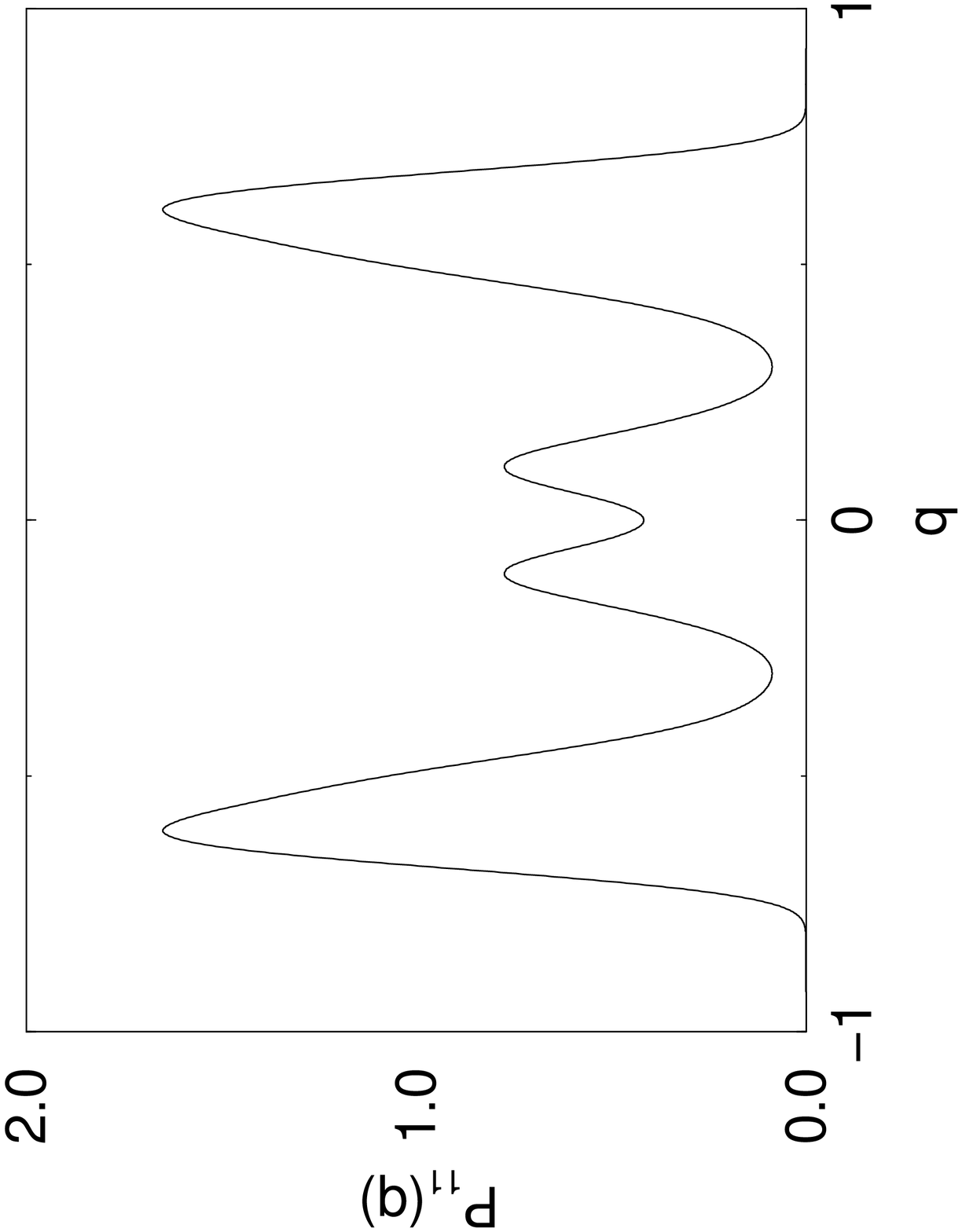}
\includegraphics{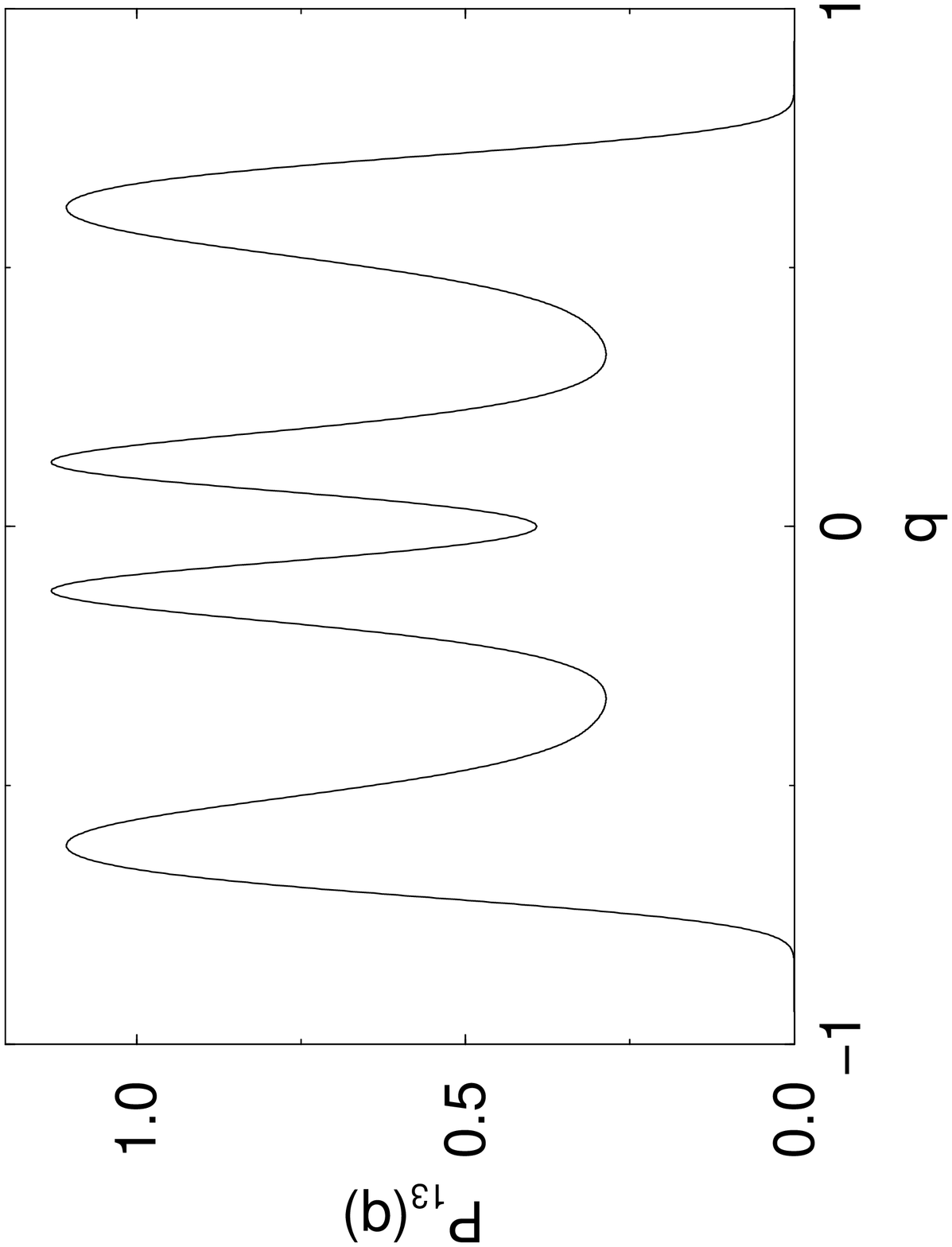}
\includegraphics{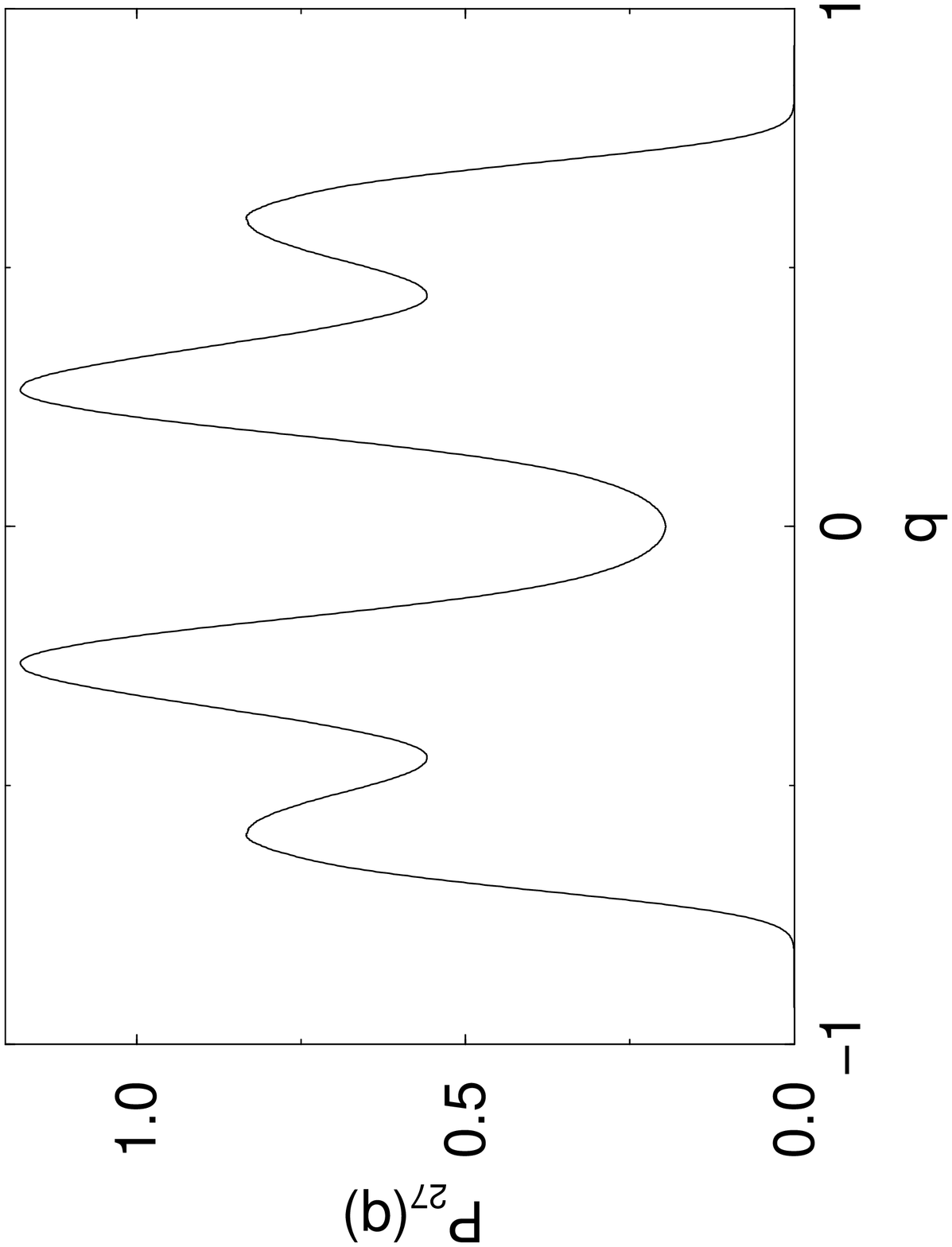}
\includegraphics{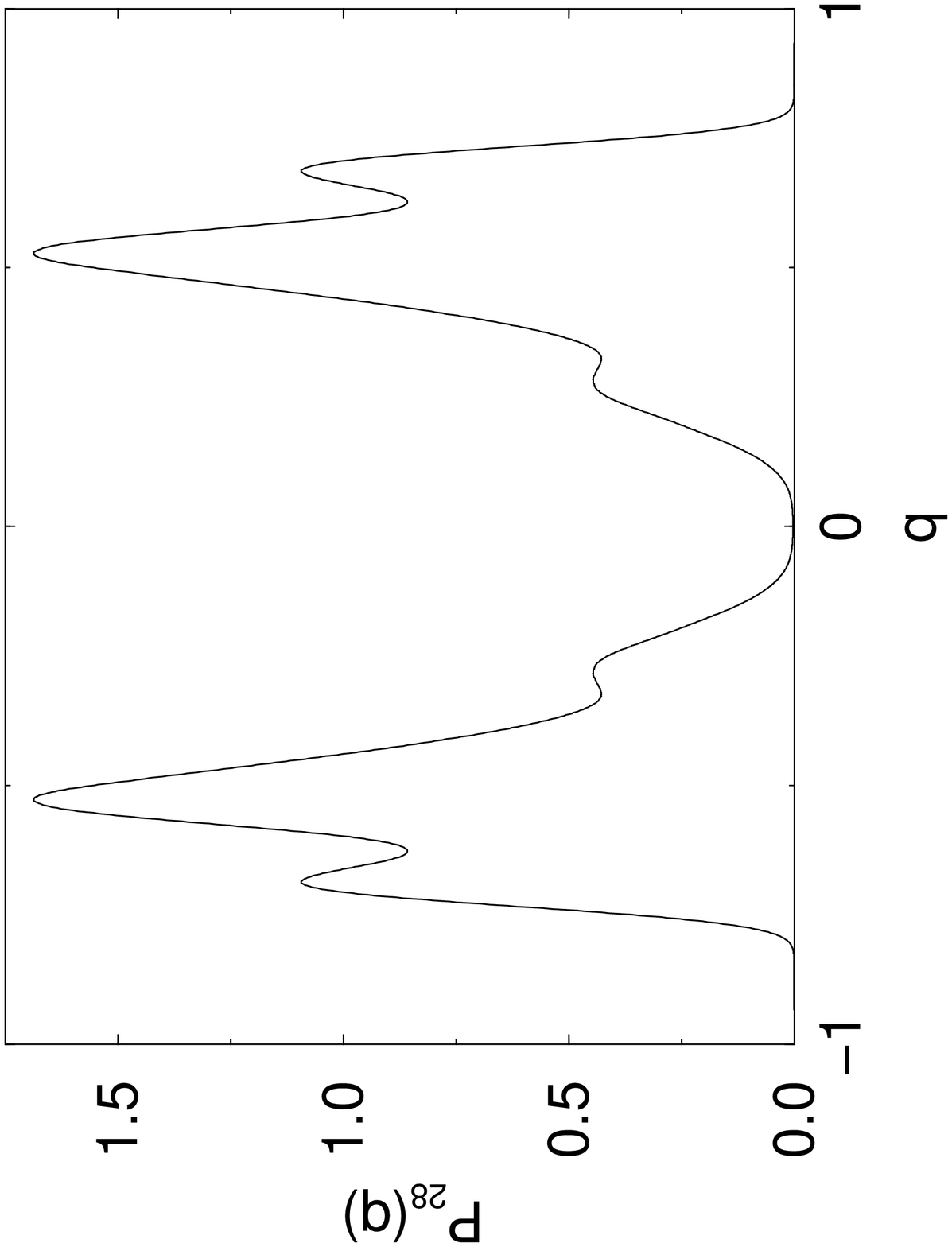}
\caption[a]{\label{fig:PJs}
Typical probability densities $P_{\cal J}(q)$ as obtained in
multi-overlap simulations of the 3D EAI spin-glass model on a $12^3$ lattice
at $T = 1 \approx 0.88 T_c$.}
\vspace*{-0.1cm}
\end{figure}

\begin{figure}[t]
\hspace*{-2cm}
\begin{center}
\vskip 6.3truecm
\includegraphics{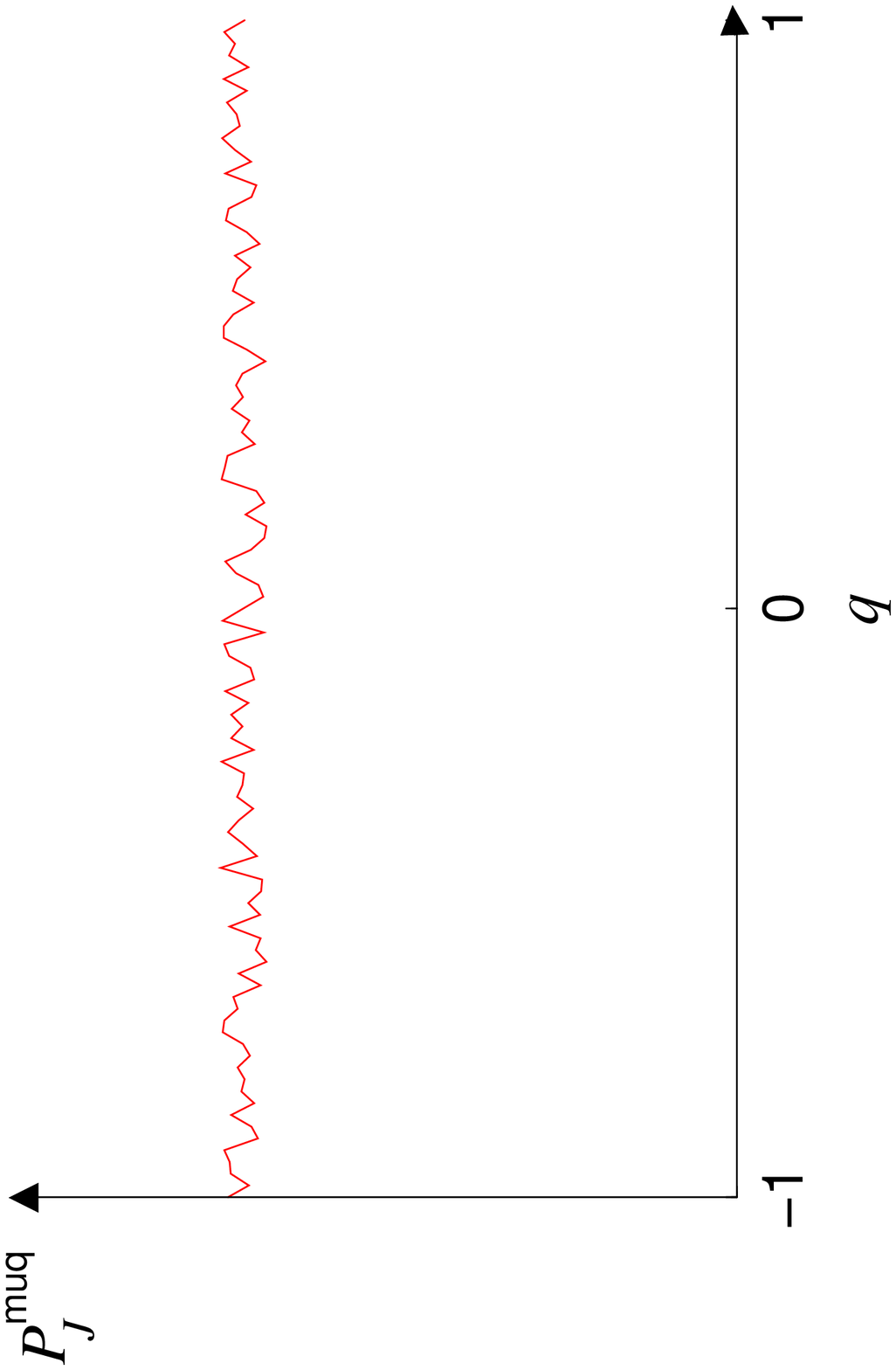}
\vspace*{-4.7cm}
{\large \hspace*{6.5cm}$$~~~~~P_{\cal J}^{\rm muq}(q) =
P_{\cal J}^{\rm can}(q) \exp[{S_{\cal J}(q)}]$$}
\vspace*{2.2cm}
\caption[a]{\label{fig:muq_sketch}
Illustration of the relation between canonical densities
$P_{\cal J}^{\rm can}(q)$ depicted in Fig.~\ref{fig:PJs} and ideally
flat multi-overlap densities $P_{\cal J}^{\rm muq}(q)$.}
\end{center}
\end{figure}

For a better understanding of 
the free-energy barriers sketched in Fig.~\ref{fig:sketch}, the probability 
densities for {\em individual\/} realizations ${\cal J}$ play the central role. 
It is, of course, impossible to get complete control over the full state space, and 
to give a well-defined meaning to ``system state'' (the $x$-axis in 
Fig.~\ref{fig:sketch}), one has to concentrate on one 
or a few characteristic properties. In our work we focused on the order
parameter and thus those free-energy 
barriers $F_B^q$ which are reflected
by the minima of $P_{\cal J}(q)$. 
A few typical shapes of $P_{\cal J}(q)$ as obtained in our simulations
are shown in Fig.~\ref{fig:PJs}.
Conventional, canonical MC
simulations are not suited for such a study because the likelihood to generate
the corresponding rare-event configurations in the Gibbs canonical ensemble is 
very small. This problem is overcome by non-Boltzmann
sampling\cite{bb_review,wj_review} with the
multi-overlap weight\cite{BeJa98}
\begin{equation} \label{eq:weight}
 w_{\cal J}(q) = \exp \left[\beta \sum_{\langle ik \rangle} 
                 J_{ik} \left(s_i^{(1)} s_k^{(1)} + s_i^{(2)} s_k^{(2)} \right) +
                 S_{\cal J}(q) \right]\ ,
\end{equation}
where the two replicas are coupled by $S_{\cal J}(q)$ in such
a way that a broad multi-overlap histogram $P_{\cal J}^{\rm muq}(q)$ over the 
entire accessible range $-1\le q\le 1$ is obtained, see Fig.~\ref{fig:muq_sketch}. 
When simulating with the multi-overlap weight (\ref{eq:weight}), canonical
expectation values of any quantity $\cal O$ can be reconstructed 
by reweighting,
$\langle {\cal O} \rangle_{\cal J}^{\rm can} = 
\langle {\cal O} e^{-S_{\cal J}} \rangle_{\cal J}/
\langle e^{-S_{\cal J}} \rangle_{\cal J}$.

The multi-overlap algorithm 
may be summarized as follows:
\begin{itemize}
\item An iterative construction of the weight
function $W_{\cal J}(q) \equiv \exp(S_{\cal J}(q))$, for each of the
quenched disorder realizations separately.
\item An equilibration period with fixed weight function.
\item The production run with fixed weight function.
\end{itemize}

Ideally $W_{\cal J}$ should satisfy $P_{\cal J}^{\rm muq}(q) =
P^{\rm can}_{\cal J}(q) W_{\cal J}(q) = {\rm const.}$, i.e., should give rise to a
completely flat multi-overlap probability density $P_{\cal J}^{\rm muq}(q)$
as sketched in Fig.~\ref{fig:muq_sketch}.
Of course, $P^{\rm can}_{\cal J}(q)$ is a priori unknown and one has to proceed 
by iteration. Let us thus assume
some approximate $W_{{\cal J},n}$ is given. The simulation would then yield
$P_{{\cal J},n}^{\rm muq}$ which, in general, is not yet perfectly flat. If
$P_{{\cal J},n}^{\rm muq}$ was sampled with arbitrary precision, the desired
weight function would
be $W_{\cal J} \propto W_{{\cal J},n}/P_{{\cal J},n}^{\rm muq}$.
For the update procedure we are actually only interested in relative
transition amplitudes and it is therefore useful to rephrase the iteration
in terms of
\begin{equation}
R_{\cal J}(q) \equiv W_{\cal J}(q+\Delta q)/
W_{\cal J}(q) = R_{{\cal J},n}(q) P_{{\cal J},n}^{\rm muq}(q)/
P_{{\cal J},n}^{\rm muq}(q+\Delta q)\ ,
\label{eq:R_recurs}
\end{equation}
where $\Delta q = 2/N$ is the step-width in $q$ as defined in Eq.~(\ref{eq:q}).
When updating the spins in the n{\em th} iteration,
$R_{{\cal J},n}(q) = W_{{\cal J},n}(q+\Delta q)/W_{{\cal J},n}(q)$
has to be considered as a given, fixed function of $q$.
The multi-overlap histogram
$P_{{\cal J},n}^{\rm muq}(q)$, however, 
is always a
noisy estimator whose statistical errors
can be estimated by $\sqrt{P_{{\cal J},n}^{\rm muq}(q)}$
(neglecting auto- and cross-correlations and assuming
unnormalized histograms counting the number of hits).
By taking the
logarithm in Eq.~(\ref{eq:R_recurs}) it is then straightforward to obtain
the squared error on $\ln R_{\cal J}(q)$,
\begin{equation}
\epsilon^2_{\ln R_{\cal J}(q)} \equiv 1/p(q) =
  1/P_{{\cal J},n}^{\rm muq}(q)
+ 1/P_{{\cal J},n}^{\rm muq}(q + \Delta q)\ .
\end{equation}
We now have two noisy estimators, $R_{{\cal J}}(q)$ and
$R_{{\cal J},n}(q)$ (with squared inverse errors $p(q)$ and $p_n(q)$),
which may be linearly combined to yield an optimized estimator
$\ln R_{{\cal J},n+1}(q) = \kappa_n(q) \ln R_{\cal J}(q) + [1-\kappa_n(q)]
\ln R_{{\cal J},n}(q)$, with
\begin{equation}
\kappa_n(q) = p(q)/[p(q)+p_n(q)]\ ,
\end{equation}
such that the statistical error of the linear combination is minimized.
By exponentiating the optimized estimator and using Eq.~(\ref{eq:R_recurs}),
we finally arrive at the recursion
\begin{eqnarray}
R_{{\cal J},n+1}(q) &=&
R_{\cal J}(q)^{\kappa_n(q)} R_{{\cal J},n}(q)^{1-\kappa_n(q)} \nonumber\\
&=&R_{{\cal J},n}(q) \left[P_{{\cal J},n}^{\rm muq}(q)/
P_{{\cal J},n}^{\rm muq}(q+\Delta q)\right]^{\kappa_n(q)},\\
p_{n+1}(q) &=& p(q) + p_n(q)\ .
\end{eqnarray}
Once $W_{\cal J}(q) = \exp(S_{\cal J}(q))$ is determined and kept fixed, the
system is equilibrated and the data production can be performed.

We measure the (pseudo-) dynamics of the multi-overlap algorithm
in terms of the autocorrelation time $\tau_{\cal J}^{\rm muq}$ which is 
defined by counting the average number of sweeps it takes to complete the
cycle $q=0 \to |q|=1 \to 0$. Adopting the usual terminology\cite{bb_review,wj_review}
for a first-order phase transition, we shall call such a cycle a ``tunnelling''
event. The weight iteration was stopped after at least  10 ``tunnelling'' events
occurred, and in the production runs we collected at least 20 ``tunnelling'' events.
To allow for standard reweighting in temperature we stored besides $P_{\cal J}(q)$
and the time series of $q$ also the energies and
magnetizations of the two replicas. The number of sweeps between measurements
was adjusted by an adaptive data compression routine to ensure that each time 
series consists of $2^{16} = 65\,536$ measurements separated by approximately 
$\tau_{\cal J}^{\rm muq}$ sweeps.

\section{Results}

Due to the large number of realizations to be simulated, the final results
are relatively costly. The studied cases are summarized in Table~\ref{tab:replicas},
where also the simulation temperatures are given: $T=1\approx 0.88 T_c$ and 
$T = 1.14 \approx T_c$ in 3D, and $T=1/0.6\approx 0.85 T_c$ in 4D. The $J_{ik}$ 
realizations were drawn using the pseudo random number generators RANMAR~\cite{Ma90}
and RANLUX~\cite{Lu94,Ja94} (luxury level 4). For the spin updates we always 
employed the faster RANMAR generator.
%
%

\begin{table}[t]

\centering
\begin{tabular}{|r|c|c|c|}
\hline
   & \multicolumn{2}{|c|}{3D} & 4D \\ \hline
$L$&$T=1.00$&$T=1.14$&$T=1/0.6$\\ \hline
 4 & 8\,192 & 8\,192 & 4\,096  \\
 6 & 8\,192 & 8\,192 & 4\,096  \\
 8 & 8\,192 & 8\,192 & 1\,024  \\
12 & ~~~640 & 1\,024 &         \\
16 &        & ~  256 &         \\
 \hline
\end{tabular}
\caption[a]{\label{tab:replicas}
Number of simulated realizations \#$\cal J$.}
\end{table}

By fitting the averaged autocorrelation times to
the power-law ansatz $\ln ([\tau_{\cal J}^{\rm muq}]_{\rm av}) = a+z\,\ln (N)$,
we obtained\cite{bbj_prb00} $z=2.32(7)$ and $z=1.94(2)$ in the 3D and 4D
spin-glass phase, respectively. Even though the quality of the fits is quite poor
and an exponential behaviour can hardly be excluded, they clearly show that the 
slowing down is quite off from the theoretical optimum $z=1$ one would expect if
the multi-overlap autocorrelation time $\tau_{\cal J}^{\rm muq}$ is dominated by 
a random-walk behaviour between $q=-1$ and $+1$.
In multicanonical simulations of the 3D model with broad {\em energy\/} histograms
an even larger exponent of $z=2.8(1)$ has been observed\cite{bhc94}.
The large values of $z$ suggest that the canonical overlap barriers are not
the exclusive cause for the slowing down of spin-glass dynamics below
the freezing point, i.e., the projection of the multi-dimensional state space
onto the $q$-direction hides important features of the free-energy landscape of 
the model.

\subsection{Free-Energy Barriers $F_B^q$}

To define effective free-energy barriers $F_B^q$ we first construct
an auxiliary 1D Metropolis-Markov chain\cite{Me53} which has the canonical 
$P_{\cal J}(q)$ probability density as its equilibrium distribution.
The tridiagonal transition matrix
\begin{equation} \label{T}
T=\left(\begin{array}{cccc}1-w_{2,1}&w_{1,2}&0 & \ldots\cr
            w_{2,1}&1-w_{1,2}-w_{3,2}&w_{2,3}&\ldots\cr
            0&w_{3,2}&1-w_{2,3}-w_{4,3} &\ldots\cr
            0 & 0 & w_{4,3} & \ldots\cr
            \vdots&\vdots&\vdots&\ddots\cr\end{array}\right)\
\end{equation}
is given in terms of the
probabilities $w_{i,j} = \frac{1}{2} \min\Bigl(1,P_{\cal J}(q_i) /
P_{\cal J}(q_j) \Bigr)$ ($ i \neq j)$
for jumps from state $q=q_j$ to $q=q_i = i/N$
in steps of $\Delta q = \pm 2/N$ or 0.
Since $T$
fulfills the detailed balance condition
(with $P_{\cal J}$)
it has only real eigenvalues.
The largest eigenvalue $\lambda_0$ equals unity and
is non-degenerate. The second largest eigenvalue
$\lambda_1$ determines the autocorrelation time of the chain (in units of
sweeps),
\begin{equation} \label{tauq}
\tau_B^q = - \frac{1}{N \ln \lambda_1} \approx \frac{1}{N (1-\lambda_1)}\ ,
\end{equation}
which we
use to define the 
associated {\em effective\/} free-energy barrier in the overlap parameter $q$ as
\begin{equation} \label{F_B^q}
 F_B^q \equiv \ln (\tau_B^q)\ .
\end{equation}

Our finite-size scaling (FSS) analyses of the thus defined overlap barriers 
are based on the (cumulative) distribution function $F(x)$. More precisely,
we constructed\cite{Be_book} a peaked distribution function $F_Q (x)$ by 
reflecting $F(x)$ at its median value 0.5,
\begin{equation} \label{F_Q}
 F_Q (x) \equiv \left\{ \begin{array}{ll}
                   F(x)\     & \quad {\rm for}\ F(x) \le 0.5 \ , \\
                   1 - F(x)\ & \quad {\rm for}\ F(x) \ge 0.5 \ . \\
                   \end{array}
                   \right .
\end{equation}
For self-averaging data
the function $F_Q$ collapses in the infinite-volume limit to
$F_Q (x) = 0.5 \ {\rm for}\ x=[x]_{\rm av}$ and 0 otherwise.
For non-self-averaging
quantities the width of $F_Q$ stays finite. The concept carries
over to quantities which diverge in the infinite-volume limit,
when for each lattice size scaled variables $x/x_{\rm med}$ are
used, where $x_{\rm med}$ denotes the median defined through
$F(x_{\rm med}) = F_Q(x_{\rm med}) = 0.5$.

\begin{figure}[tb]
\hbox{\psfig{figure=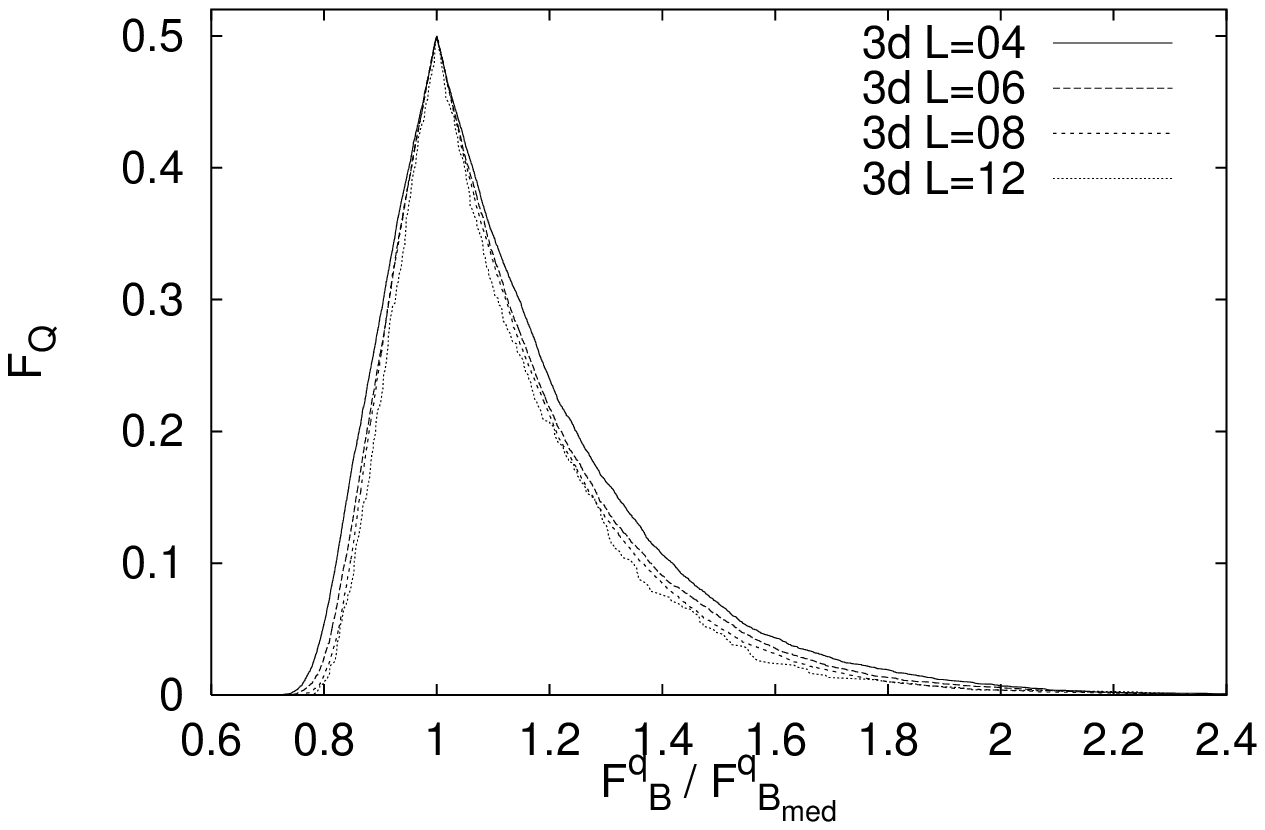,width=6.3cm}
\psfig{figure=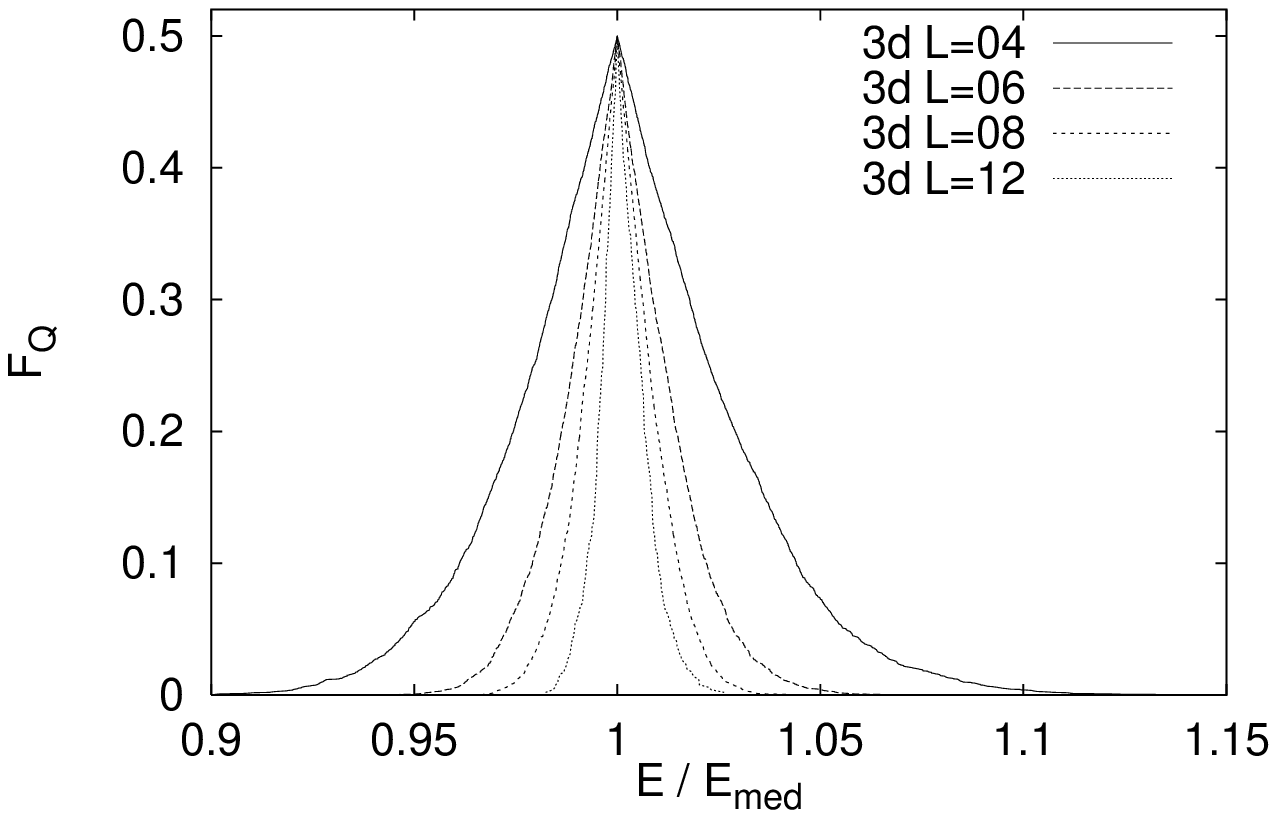,width=6.3cm} }
\caption{\label{fig:FqB3d}
Distribution function $F_Q$~(\ref{F_Q}) for the 3D overlap
barriers~(\ref{F_B^q}) (left) and the energy (right) in the spin-glass phase
at $T=1$ in units of their median values.
} 
\end{figure}

The behaviour of $F_Q(F_B^q/F^q_{B_{\rm med}})$
shown on the l.h.s.\ of Fig.~\ref{fig:FqB3d} for the 3D case at $T = 1$ clearly
suggests that $F^q_B$ is a non-self-averaging quantity. In
4D the evidence is even stronger than in 3D.
Non-self-averaging was also observed\cite{bbj_prb00} for the
autocorrelation times $\tau_{\cal J}^{\rm muq}$ of our algorithm while
the energy is an example for a self-averaging quantity; cf.\ the r.h.s.\ of
Fig.~\ref{fig:FqB3d}.
For non-self-averaging quantities
one has to investigate many samples and should report
the FSS behaviour for fixed values of the
cumulative distribution function $F$.

\begin{figure}[tb] 
\hbox{\psfig{figure=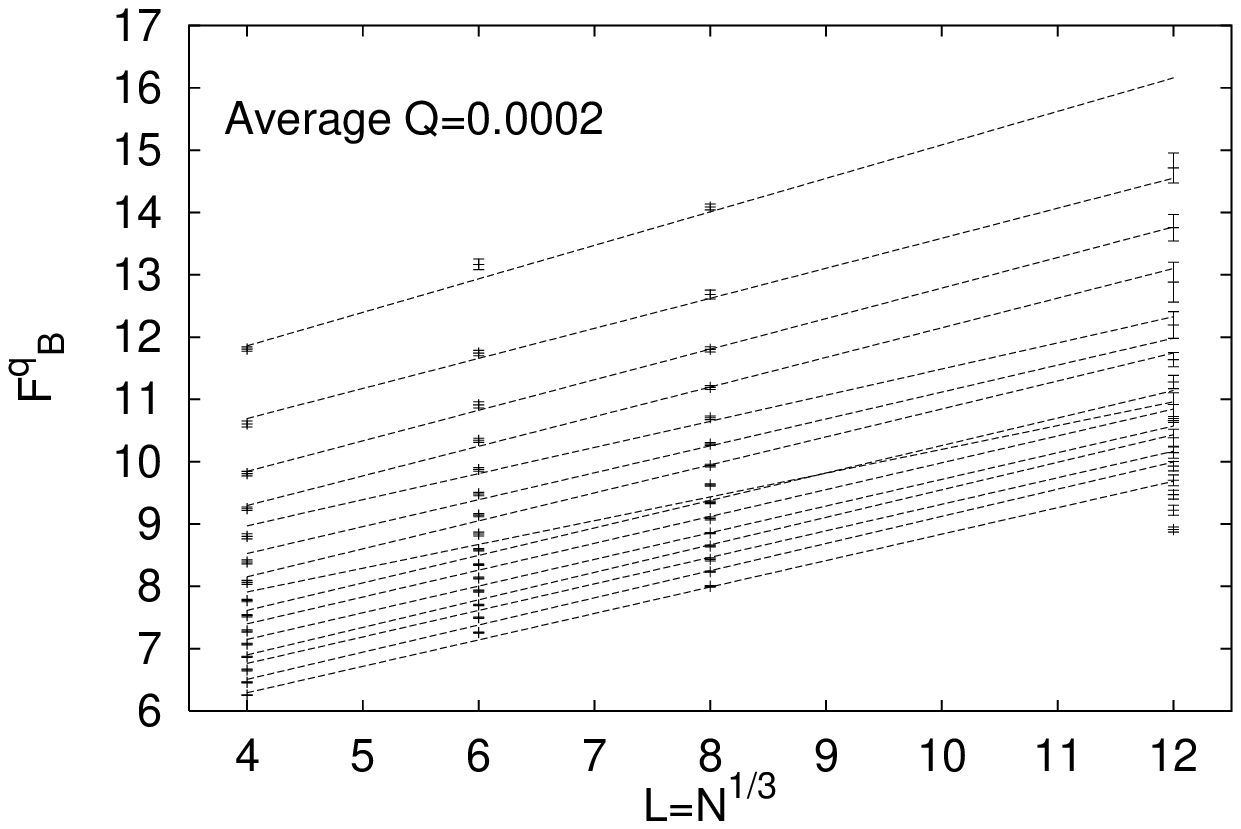,width=6.4cm}
\psfig{figure=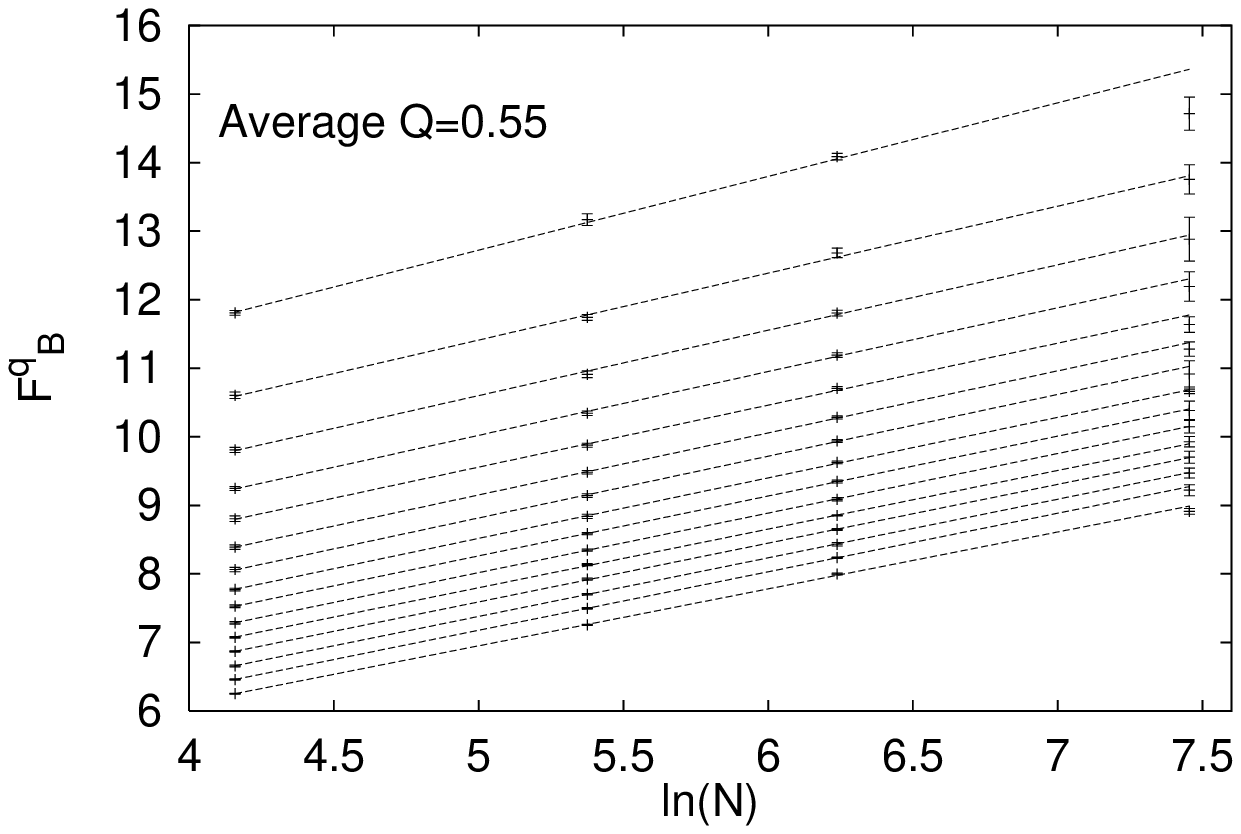,width=6.4cm} }
\caption{\label{fig:all_fits}
FSS fits of the 3D overlap barriers $F_B^q$ in the spin-glass phase
at $T=1$ for fixed values of the distribution function, 
$F = i/16$, $i=1,\dots,15$ (from bottom to top). Shown are the results
for the mean-field prediction (\ref{F_Bfits1})
(left) and the logarithmic ansatz (\ref{F_Bfits2}) (right).
}
\end{figure}
%

In Ref.~15 we performed FSS fits for $F=i/16$, $i=1,\dots ,15$,
assuming an ansatz suggested by mean-field theory\cite{MaYo82,unpublished},
\begin{equation} \label{F_Bfits1}
   F_B^q = a_1 + a_2\, N^{1/3}\ ,
\end{equation}
corresponding to $\tau_B^q \propto e^{a_2\, N^{1/3}}$.
In both dimensions the goodness-of-fit parameter\cite{NumRec} $Q$
turned out to be unacceptably small.
The 3D fits are depicted on the l.h.s.\ of Fig.~\ref{fig:all_fits}.
We therefore also tried fits to the ansatz
\begin{equation} \label{F_Bfits2}
   F_B^q = \ln (c) + \alpha\, \ln (N)\ ,
\end{equation}
corresponding to
$\tau_B^q \propto N^{\alpha}$; cf.\ the r.h.s.\ of Fig.~\ref{fig:all_fits}.
Since in 3D as well as in 4D the average $Q$-value is now within
the statistical expectation, the latter ansatz (\ref{F_Bfits2}) is strongly
favoured over the mean-field prediction (\ref{F_Bfits1}).
As a function of $F$ ($ = 1/16$ -- $15/16$) the exponent
$\alpha=\alpha (F)$ in the  power law (\ref{F_Bfits2}) varies smoothly
from 0.8 to 1.1 in 3D and from 0.8 to 1.3 in 4D. A similar
analysis\cite{bbj_prb00} for the autocorrelation times $\tau_{\cal J}^{\rm muq}$ of
the multi-overlap algorithm gives exponents $\alpha (F)$ which are
larger, $\alpha^{\rm muq} (F) \approx \alpha^q_B (F) +1$. This
is in agreement with our
observation that other relevant barriers exist,
which cannot be detected in the overlap parameter $q$.

\subsection{Averaged Probability Densities $P(q)$}
\begin{figure}[-t] \begin{center}
\hbox{\epsfig{figure=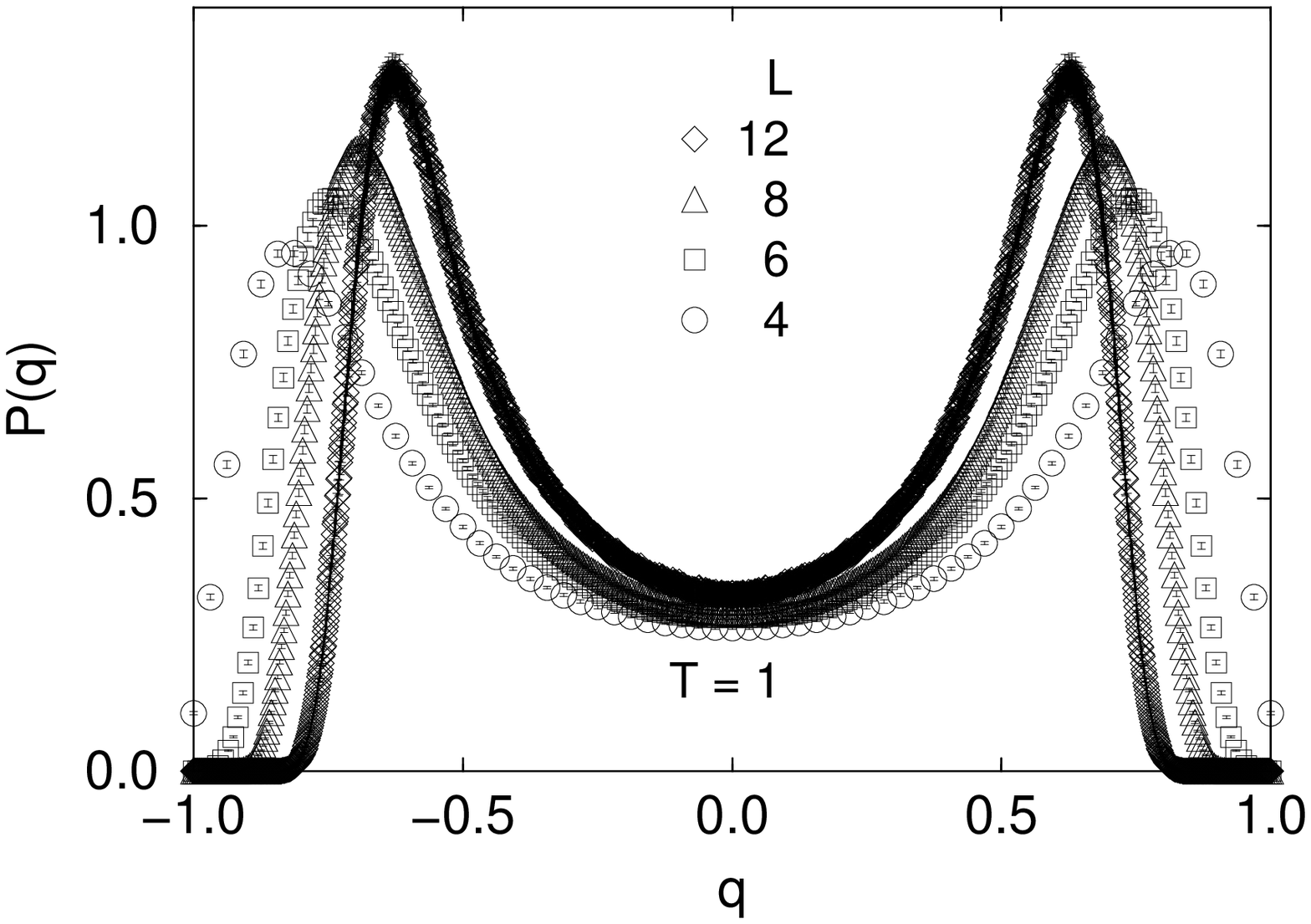,width=6.25cm}
\epsfig{figure=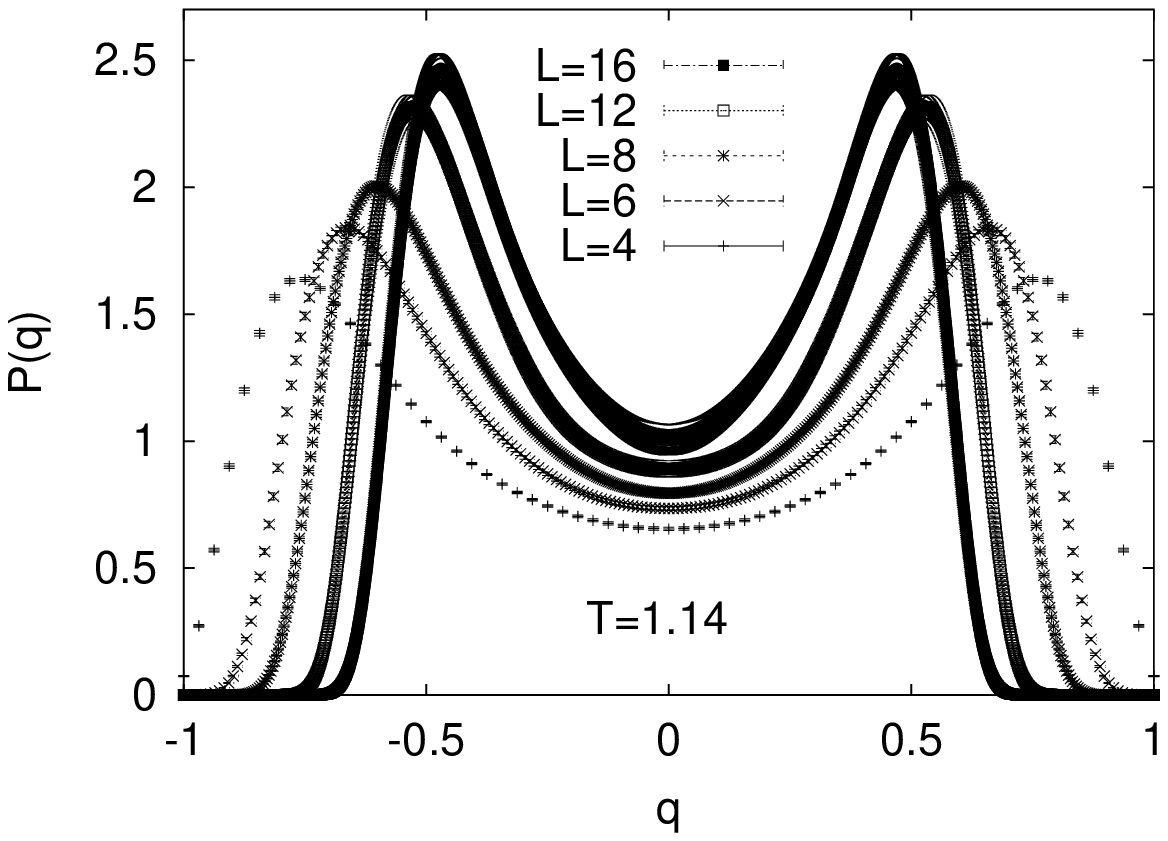,width=6.65cm}}\vspace*{-0.5cm}
\end{center} 
\caption[a]{\label{fig:P_q}
Overlap probability densities for the 3D EAI model in the spin-glass
phase (left) and close to criticality (right).
}
\end{figure}

The averaged canonical densities $P(q)$ of the 3D model are
shown in Fig.~\ref{fig:P_q} for both $T = 1 \approx 0.88 T_c$ and
$T=1.14 \approx T_c$. At least close to $T_c$ one expects that, up to
finite-size corrections, the probability densities scale with system
size. A method to confirm this visually is to plot $P'(q) \equiv \sigma P(q)$
versus $q' = q/\sigma$, where $\sigma$ is the standard deviation. By fitting the 
standard deviation to the expected FSS form $\sigma = c_1 L^{-\beta/\nu}$ we
obtained\cite{bbj_pre01},
\begin{eqnarray} \label{beta_nu1}
 \frac{\beta}{\nu} &=& 0.312(4)\ ,\ \, Q=0.32\ {\rm ~for}\ T=1.14\,,
 ~~{\rm and} \\ \label{beta_nu2}
 \frac{\beta}{\nu} &=& 0.230(4)\ ,\ \, Q=0.99\ {\rm ~for}\ T=1\,.
\end{eqnarray}
On the l.h.s.\ of Fig.~\ref{fig:Pp3d} we show the
scaling plot\cite{bbj_pre01} for $T = 1.14$ which 
demonstrates that
the five probability densities collapse onto a single master curve.

\subsection{Tails of $P(q)$}

\begin{figure}[-t] \begin{center} 
\hbox{\epsfig{figure=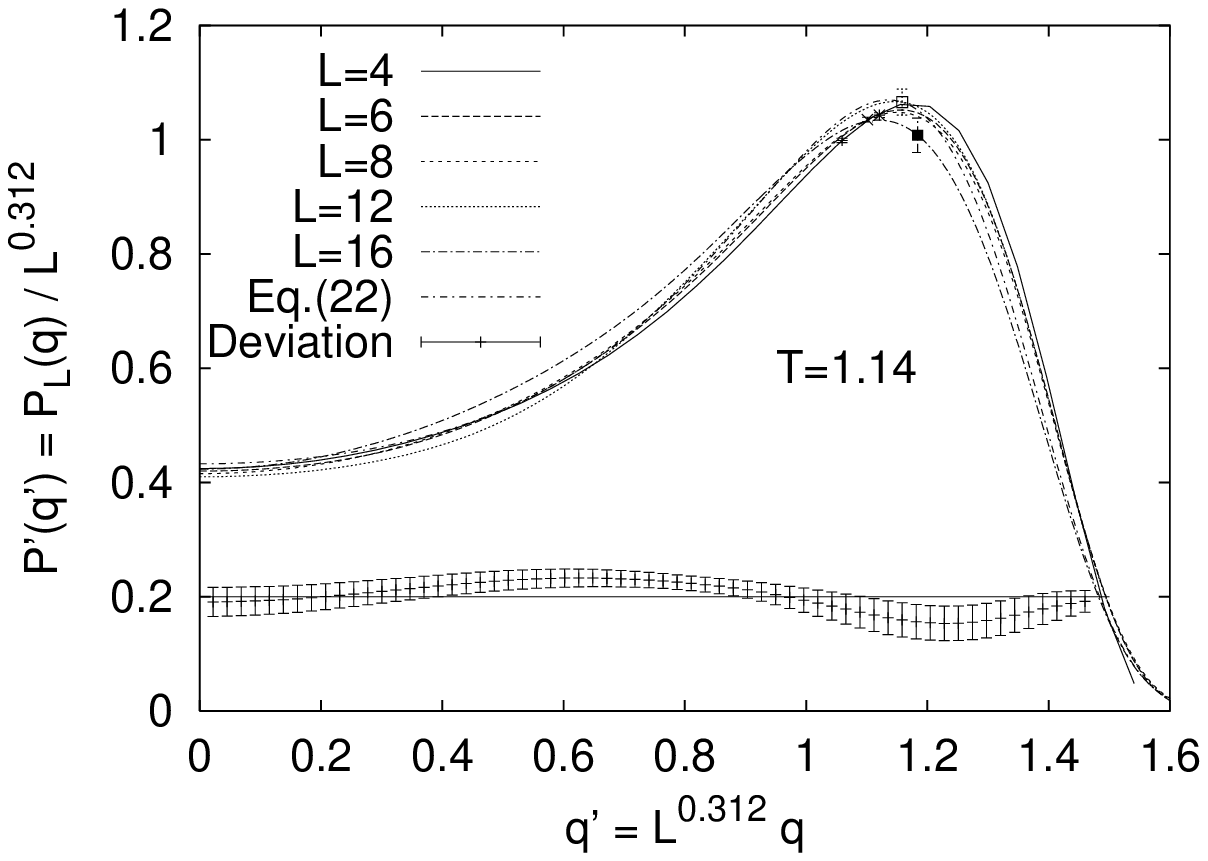,width=6.5cm}
\epsfig{figure=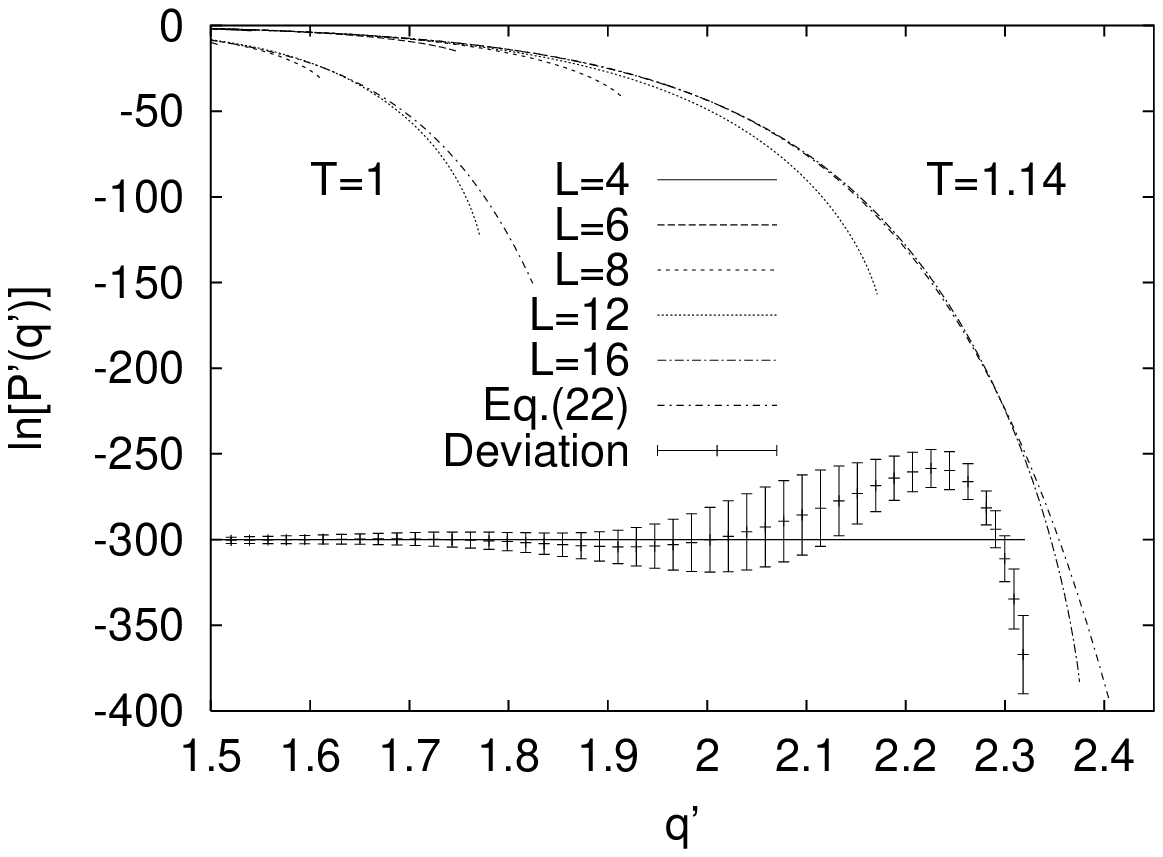,width=6.5cm}}
\caption{\label{fig:Pp3d}
Rescaled overlap probability densities for the EAI spin-glass model
on $L^3$ lattices at the critical temperature on a linear (left) and
expanded logarithmic (right) scale. In the lower part of the plots
the deviation $ P'_{16}(q') - P'_{\rm fit}(q') \pm \triangle
P'_{16}(q')$ of some $L=16$ data from the modified Gumbel fit (\ref{eq:mod_Gumbel})
is shown  (with an unimportant offset added in order to fit inside the figure).
}
\end{center} \end{figure}

The multi-overlap algorithm becomes particularly powerful when
studying the tails of the probability densities which are highly suppressed
compared to the peak values; see the r.h.s.\ of Fig.~\ref{fig:Pp3d}
which shows $P(q)$ at $T=1.14$ over more than 150 orders of magnitude.
Based on the replica mean-field approach,\, theoretical\, predictions\, for
the scaling behaviour of the tails have been
derived by Parisi and collaborators\cite{Pa92}. They showed that
$P(q) = P_{\max}\, f(N\,(q-q_{\max}^{\infty})^x)$
for $q > q_{\max}^{\infty}$ and concluded more quantitatively that
\begin{equation} \label{Pq_tail2}
P(q) \sim \exp \left[ - c_1\, N\, (q-q_{\max}^{\infty})^x \right]
 ~~{\rm for}~~  N\,(q-q_{\max}^{\infty})^x ~~{\rm large}\ ,
\end{equation}
with a mean-field exponent of $x=3$.
By allowing for an overall
normalization factor $c_0^{(N)}$ and taking the logarithm twice we
have performed fits of the form\cite{china,dubna}
\begin{equation} \label{lnln_tail_fit}
Y \equiv \ln \left[ -\ln (P/c_0^{(N)}) \right] - \ln N =
\ln c_1 + x\, \ln (q-q_{\max}^{\infty})\ .
\end{equation}
Leaving the exponent $x$ as a free parameter, we arrived at the
estimates
$x = 12(2)$ in 3D at $T = 1$ and
$x = 5.3(3)$ in 4D at $T = 1/0.6$
which are both much larger than the mean-field value of $x=3$.

By looking for reasonable alternatives we realized that for many other
systems the {\em statistics of extremes\/} as pioneered by Fr\'echet, 
Fisher and Tippert, and von Mises\cite{Gu58,Ga87} has let to a good 
ansatz with universal properties\cite{Br00,Br01}.
It is based on a standard result\cite{Gu58,Ga87},
due to Fisher and Tippert, Kawata, and Smirnow, for the universal
distribution of the first, second, third, $\dots$ smallest of a set of
$N$ independent identically distributed random numbers. For an
appropriate, exponential decay of the random number distribution
their probability densities are given by the Gumbel form
\begin{equation} \label{eq:Gumbel}
f_a(x)\ =\ C_a\,\exp\left[\,a\,\left( x-e^x\,\right)\right]\ ,
\end{equation}
in the limit of large $N$.
The exponent $a$ takes the
values $a=1,2,3,\dots $, corresponding, respectively, to the first,
second, third, $\dots$ smallest random number of the set,
$x$ is a scaling variable which
shifts the maximum value of the probability density to zero, and
$C_a$ is a normalization constant. For certain spin-glass systems the
possible relevance of this universal distribution has been pointed out
by Bouchaud and M\'ezard\cite{BoMe97}, and 
for instance also in applications to the 2D XY
model in the spin-wave approximation\cite{Br00,Br01}
the Gumbel ansatz (\ref{eq:Gumbel})
fitted well, albeit with a modified value of $a = \pi/2$. 

In our case we set $x = b (q' - q'_{\rm max})$ and modified the first
$x$ on the r.h.s.\ of (\ref{eq:Gumbel}) to $c \tanh(x/c)$, where $c>0$ is
a constant, in order to reproduce the flattening of the densities towards
$q'=0$.	Notice that in the tails of the densities, i.e. for large, positive $x$, 
this term is anyway subleading. A symmetric expression for $P'(q')$ reflecting
its $q' \to -q'$ invariance is obtained by multiplying the above
construction with its reflection about the $q'=0$ axis,
\begin{equation} \label{eq:mod_Gumbel}
P'(q') =  C \exp \left\{a \left[c \tanh \left(\frac{b}{c}\,(q'-q'_{\max})
\right) - e^{+b\,(q'-q'_{\max})}\right]\right\} \times (q' \to -q')\ .
\end{equation}  
Of course, the important large-$x$ behaviour of Eq.~(\ref{eq:Gumbel})  
is not at all affected by our manipulations.

By fitting this ansatz to our 3D data we obtained final estimates
of $a=0.448(40)$ for $T=1.14$ and $a=0.446(37)$ for $T=1$, respectively. For
$T=1.14$ our best fit is already included on the l.h.s.\ of Fig.~\ref{fig:Pp3d}.
We see a good consistency between the data and the fit over a remarkably wide range
of $q'$. Even more impressive is the excellent agreement in the tails of the 
densities. Taking the $T=1.14$, $L=16$ result at face value, we find\cite{bbj_pre01}
a very good fit over a remarkable range of $200/\ln(10) \approx 87$ orders of 
magnitude!

\section{Summary and Conclusions} \label{sec_conclude}

Employing non-Boltzmann sampling with the multi-overlap MC
algorithm we have investigated the probability densities $P_{\cal J}(q)$
of the Parisi order parameter $q$.
The free-energy barriers $F_B^q$ as defined in Eq.~(\ref{F_B^q}) turn out 
to be non-self-averaging. The logarithmic scaling ansatz (\ref{F_Bfits2})
for the barriers at fixed values $F$ of their cumulative distribution function
is found to be favoured over the mean-field
ansatz (\ref{F_Bfits1}). Further, relevant barriers are still reflected in
the autocorrelations of the multi-overlap algorithm.

The averaged densities exhibit a pronounced FSS collapse onto a common
master curve even in the spin-glass phase. For the scaling of their
tails towards $q = \pm 1$ we find qualitative agreement with the decay
law predicted by mean-field theory, but with an exponent $x$ that is,
in particular in 3D, much larger than theoretically expected. A much
better fit over more than 80 orders of magnitude is obtained in 3D by using a
modified Gumbel ansatz, rooted in extreme order statistics\cite{Gu58,Ga87}. 
The detailed relationship between the EAI spin-glass model and extreme order 
statistics remains to be investigated, and it is certainly a challenge to
extend the work of Bouchaud and M\'ezard\cite{BoMe97} to the more
involved scenarios of the replica theory.

\section*{Acknowledgements}
We would like to thank A. Aharony, K. Binder, F. David, E. Domany and A. Morel
for useful discussions. The project was partially supported by the 
German-Israel-Foundation (GIF-I-653-181.14/1999), the US Department of Energy 
(DOE-DE-FG02-97ER41022), and the T3E computer-time grants hmz091 (NIC, J\"ulich), 
p526 (CEA, Grenoble) and (in an early stage of the project) bvpl01 (ZIB, Berlin).


\bibliographystyle{unsrt}

\end{document}